\documentclass[twoside]{dis08}
\usepackage[latin1]{inputenc}
\usepackage[dvips]{graphicx,epsfig,color}
\usepackage{wrapfig,rotating}
\usepackage{amssymb,amsmath,array}

\begin{document}
\title{Electroweak Physics Measurements at the LHC}

\author{
N. Besson
%
\thanks{On behalf of the ATLAS and CMS collaborations.}
%
\vspace{.3cm}\\
%
DSM/Irfu/SPP, CEA/Saclay,\\
F-91191 Gif-sur-Yvette Cedex
%
}

\maketitle

\begin{abstract}
Although often quoted as a discovery collider, the LHC will also allow for precise measurements. 
In particular, in the electroweak sector, the determination of the masses of the top quark 
and the $W$ boson will benefit from high statistics and new methods~\cite{slides}.
\end{abstract}

\section{Introduction}

In the Standard Model, the electroweak observables are related through radiative
corrections which allow to put indirect constraints on the mass of the Standard Model
Higgs boson, $M_H$, with an uncertainty currently mainly dominated by the uncertainties on 
the top quark and $W$ boson masses, $M_t$ and $M_W$.
With $M_t=172.6\pm 1.4$~GeV~\cite{top_TeV} and 
$M_W=80.398\pm 0.025$~GeV~\cite{W_TeV}, the most recent Standard
Model global fit gives a preferred value of 
$M_H = 87^{+36}_{-27}$~GeV and
an upper limit of 160~GeV~\cite{lepewwg}.

The LHC will deliver around 
$8\times 10^5$ $t\bar{t}$ pairs and $20\times 10^6$ $W$ bosons (per leptonic
decay channel) per fb$^{-1}$. 
Thus statistical uncertainties  
will become negligible very fast and the real
challenge will be to control the systematic uncertainties of the mass measurements.

In the next sections, I will present examples of $M_t$ (Sec.~\ref{top-mass}) and $M_W$
(Sec.~\ref{W-mass}) measurements by the ATLAS and CMS collaborations and what 
kind of accuracies we can expect.

\section{Measurements of the mass of the top quark at LHC}
\label{top-mass}

\subsection{Top quark pairs at LHC}

At the LHC, top quark pairs are mainly
produced via gluon fusion, yielding a production cross-section of 833~pb at next to
leading order: LHC will be a top factory.

The most favorable channel is the semi-leptonic channel 
($tt \rightarrow WbWb \rightarrow (\ell\nu) b + (\mbox{jj})b$) where the full 
topology of the final state is used to select the signal with good purity and the 
hadronic side is used to measure the top quark mass.
Analysis of the fully hadronic and dileptonic channels are described here~\cite{Davids:962027}.

\subsection{Top quark mass in the semi-leptonic channel using the mass peak}

A first analysis is based on sequential selection cuts followed by a fit of the
mass peak~\cite{ATLAS:1103727}.

The event selection requires one high-$p_T$ isolated lepton inside 
the acceptance, high
missing transverse energy
and at least 4 high-$p_T$ jets, among which 2 are tagged as $b$-jets.
This ensures a signal over background ratio of the order of 10. The main backgrounds
are single top events, mainly reduced by the 4 jets cut, fully hadronic
$t\bar{t}$ events, reduced by the cuts on leptons, $W$+jet 
and $Z$+jet events. Backgrounds from QCD multi-jet events and $b\bar{b}$ production are
negligible after leptonic cuts and backgrounds from di-boson events have a much smaller
contribution and are strongly reduced by the cuts on jets.

To reconstruct the hadronic side of the decay, an {\it{in situ}} rescaling is performed
by minimizing $\chi^2 = \frac{(M_{jj}-M_W^{PDG})^2}{\Gamma^2_W} 
       + \frac{(E_{j_1}(1-\alpha_1))^2}{\sigma^2_1} 
       + \frac{(E_{j_2}(1-\alpha_2))^2}{\sigma^2_2}$.
The minimization constrains the light jet pair mass $M_{jj}$ to $M_W$,
via corrections $\alpha_i$ to the light jet energies $E_{j_i}$.
All possible jet combinations are tried; the one minimizing  the $\chi^2$ is kept.
The $b$-jet closest to the hadronic $W$ boson is associated to the chosen pair. 
The 3 jets invariant
mass is then fitted with a Gaussian (a polynomial is added to account for the
mainly combinatorial remaining background).

\begin{wraptable}{l}{0.5\columnwidth}
\centerline{\begin{tabular}{|l|r|}
\hline
Source & Effect on $M_t$  \\\hline
Light jet energy scale    & 0.2~GeV/\%  \\
$b$-jet energy scale      & 0.7~GeV/\%  \\
ISR/FSR                   & $\simeq$ 0.3~GeV/\%  \\
$b$ fragmentation         & $\leq$ 0.1~GeV/\%  \\
Background                & negligible  \\
\hline
\end{tabular}}
\caption{Sources of systematic uncertainties and their effect on $M_t$.}
\label{top_sys_1}
\end{wraptable}

The most important sources of systematic uncertainties are listed in table~\ref{top_sys_1}.
The main contribution is the jet energy scale (JES) on which $M_t$ depends linearly, 
with a slope of only 0.2~GeV/\% (light jets), thanks to the {\it{in situ}} rescaling. 
For this analysis, the JES was taken from fits of the difference between 
the reconstructed and generated jet energies. It can also be measured 
on data with a
template method. This was studied and all systematic uncertainties were found below 0.5~\% 
yielding an uncertainty on the JES of 1~\% per fb$^{-1}$.
The result is $M_t = 175.0 \pm 0.2 (stat.) \pm 1.0 (syst.)$~GeV, for an 
input mass of 175~GeV and 1~fb$^{-1}$.

\subsection{Top quark mass in the semi-leptonic channel using the most probable mass}

Another analysis of the semi-leptonic channel relies on likelihoods and the choice of
the most probable mass~\cite{D'Hondt:951376}.

\begin{wrapfigure}{r}{0.5\columnwidth}
\centerline{\includegraphics[width=0.45\columnwidth]{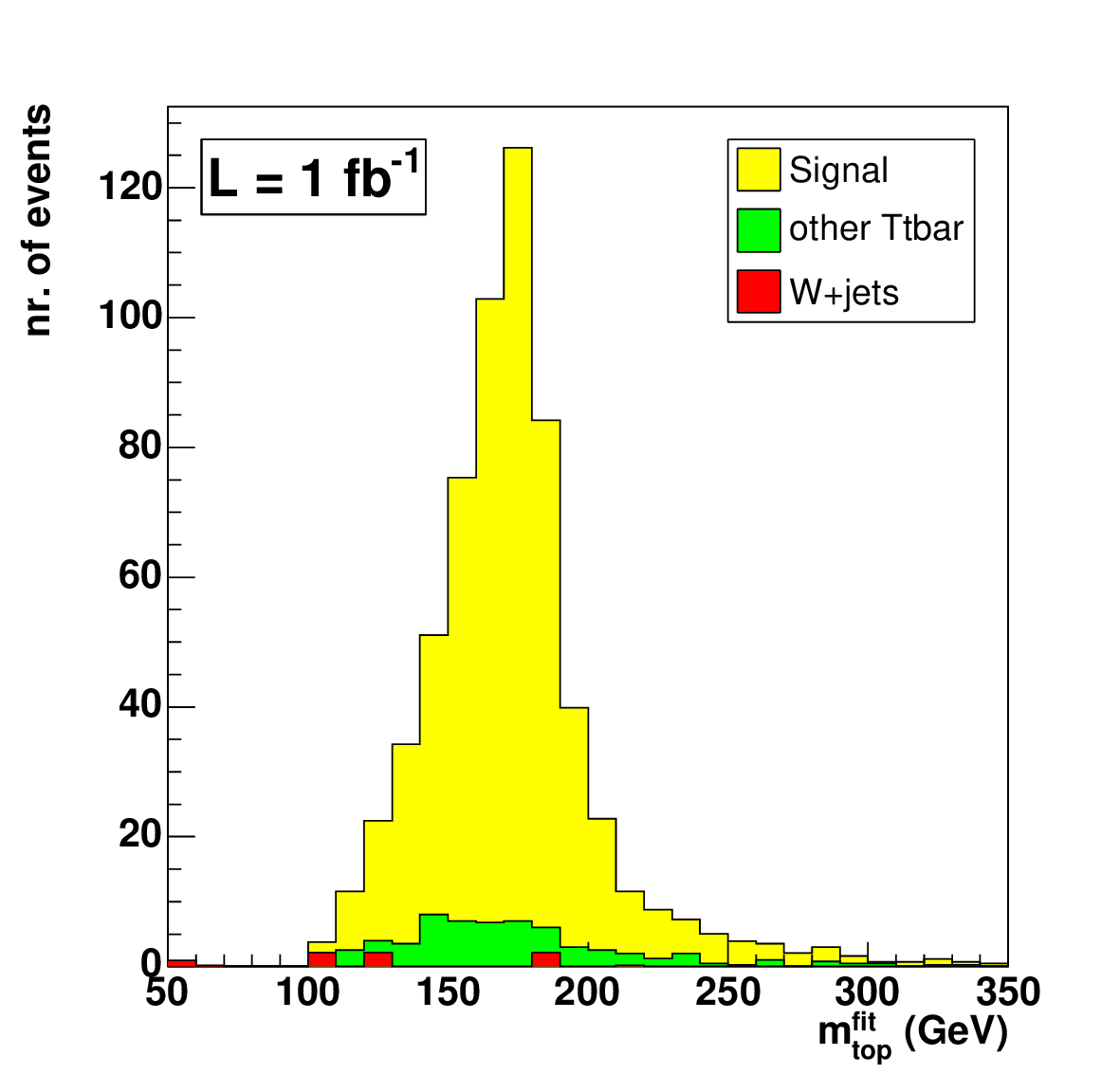}}
\caption{Distribution of the mass of the hadronic decaying top quark
after applying the kinematic fit.}\label{CMS_top}
\end{wrapfigure}

A first likelihood, taking into account the probability for leptons to come from $W$ 
boson decays and the $p_T$ of the jets and leptons, gives the probability for an event to be 
a signal event. A second probability for a combination to be the right one is based on angles 
between jets and between jets and leptons. 
Then a kinematic fit is applied which adapts 
the jet momenta to agree with the world average value of $M_W$ (Figure~\ref{CMS_top}). 
In the fit, one forces the value of $M_t$ scanning the 
whole mass range and obtains an explicit likelihood as a function of $M_t$ for each event. 
The likelihoods for all events are combined and the maximum 
likelihood technique gives the estimator for $M_t$.

For 1~fb$^{-1}$ and
in the $\mu$ channel only, the mass scan method gives a statistical uncertainty of 0.66~GeV 
and a systematic uncertainty of 1.13~GeV dominated here again by the JES. 

\section{$W$ boson mass measurements}
\label{W-mass}

\subsection{Method}

To measure $M_W$, only $W\rightarrow \ell \nu$ final states are useful. Because of the neutrino
other observables than $M_W$ have to be used: for instance $p_T^\ell$,
 which shows a Jacobian peak at $M_W/2$.
The template method is used: the reconstructed 
distribution is tested, with a $\chi^2$ test, against a set of
template distributions, characterized by a mass scale $\alpha_M$. 
The minimum of the parabola describing the result of the tests 
{\it vs} $\alpha_M$ gives $M_W$.
This relies crucially on the control of any effect distorting the $p_T^\ell$
distribution.
These effects come from different sources: experimental 
(lepton energy scale, linearity and resolution, $E^\ell$ non-Gaussian tails,
 efficiencies), theoretical 
(initial and final state radiation, boson width, PDF) and environmental 
(backgrounds, underlying event, pile-up).
Most of them can be strongly constrained with $Z$ measurements.

\subsection{Creation of templates from $Z$ events}

In the scaled observable method~\cite{Buge:951371}, the template distributions are 
created by transforming the $Z$ distributions into $W$ ones.
The distributions from the $Z$ events (e.g. $p_T^\ell$)
are scaled to the $W$ mass:
after selection, a lepton is randomly chosen and removed from the $Z\rightarrow \ell\ell$ 
events. 
Then the chosen lepton spectrum is scaled from the $Z$ to the $W$ mass. The scaled distributions 
are weighted with 
the ratio of the differential cross-sections, including
theoretical calculations and simulation to account for differences between
  $W$ and $Z$ in acceptance and detector effects.  
Most common experimental and theoretical uncertainties cancel.
The $\chi^2$ test is made between the scaled spectra. 
The analysis of the electron channel with the $p_T^e$ distribution and for 1 fb$^{-1}$ 
gives a statistical uncertainty of 40 MeV.
The experimental uncertainties, 40 MeV, are dominated by the lepton 
 energy scale linearity. The theoretical uncertainties are estimated to be 40 
 MeV and are dominated by the PDF uncertainties.

\subsection{Calibration of templates with $Z$ constraints}

\begin{wrapfigure}{r}{0.5\columnwidth}
\centerline{\includegraphics[width=0.45\columnwidth]{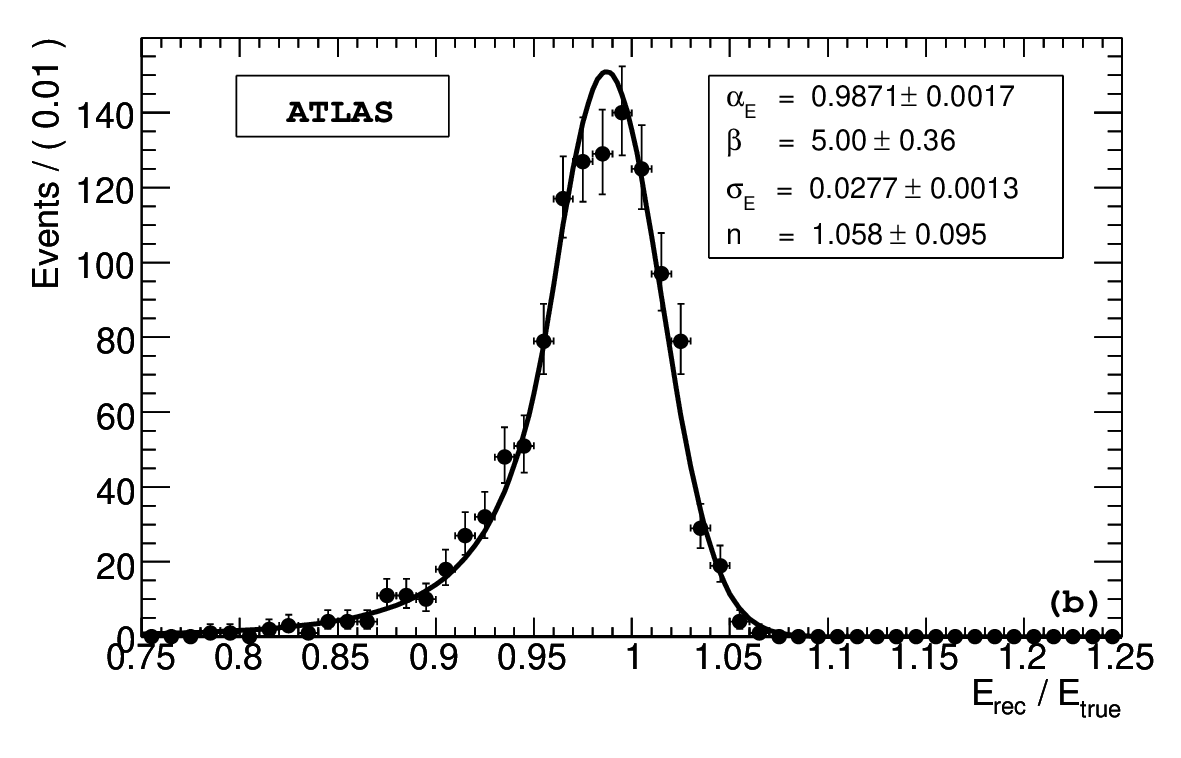}}
\caption{Example of reconstructed over simulated electron energy with non-Gaussian tails.}\label{ngt}
\end{wrapfigure}
Template distributions can also be calibrated with 
$Z$ events~\cite{Z_paper}. The effects are measured on $Z$ events and
 propagated to the $W$ sample.
As a demonstration, an exercise is made with the lepton $p_T$ distribution 
in the $W\rightarrow e\nu$ channel and for 15 pb$^{-1}$~\cite{W_CSC}.
Effects like non-Gaussian tails (see example on Figure~\ref{ngt}) are taken into account in the determination
with $Z\rightarrow ee$ events of the energy scale and 
resolution. The results are fed back to the $W$ mass fit.
Templates are created with generated electrons from $W\rightarrow e\nu$, smeared with 
the parameters found on the $Z$ events. The result shows no bias and the global uncertainty is 
$\delta M_W = 110 (stat) \pm 114 (exp.) \pm 25 (PDF)$~MeV.
The main source of uncertainty is the lepton energy scale. 

\begin{wrapfigure}{r}{0.5\columnwidth}
\centerline{\includegraphics[width=0.45\columnwidth]{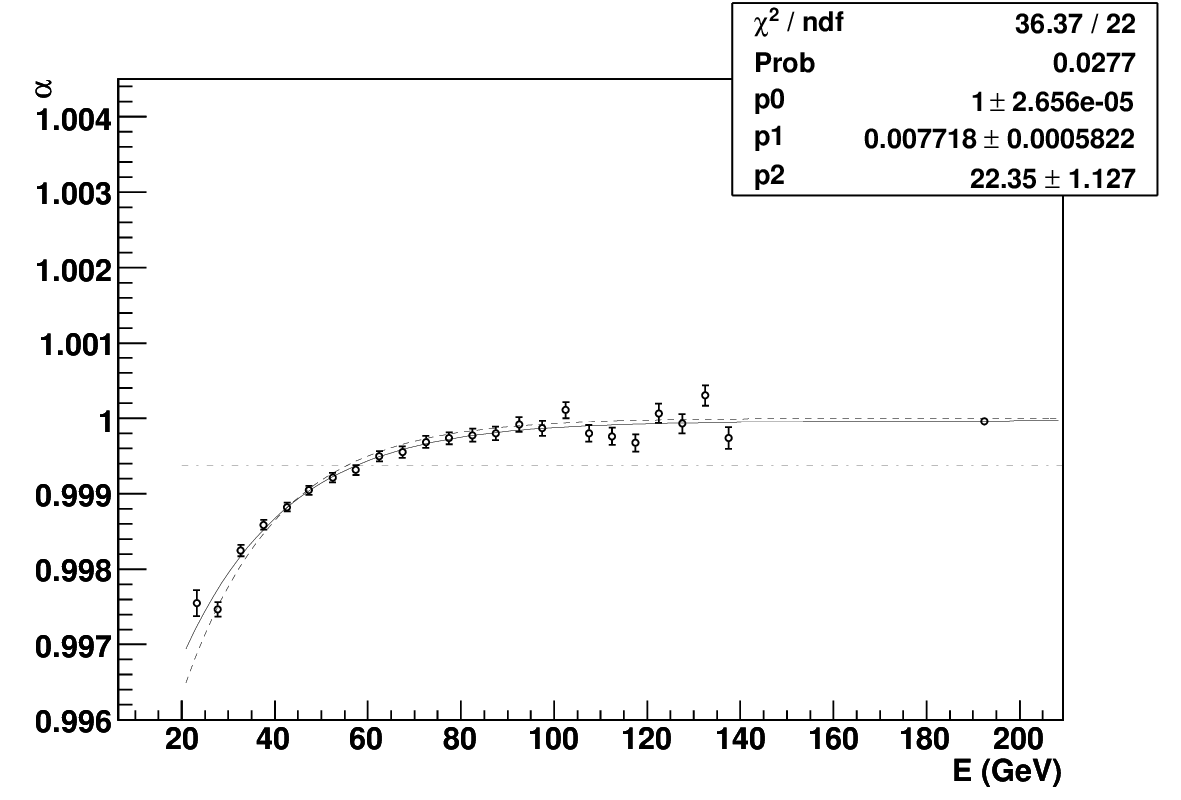}}
\caption{Energy dependent detector scale, as determined from fits to the 
$Z\rightarrow ee$ mass peak.}\label{lin}
\end{wrapfigure}

In the long term~\cite{W_paper}, with a luminosity of 10~fb$^{-1}$, the energy scale can be determined as above
 and in bins of lepton $p_T$, allowing to control the detector linearity
 up to $2\cdot 10^{-4}$ (Figure~\ref{lin}).
 
This reduces the uncertainty on $M_W$ to 4~MeV. The same can be done for resolution. 
The remaining uncertainty is 1~MeV.

With high statistics the impact of the W 
$p_T$ and rapidity on the mass fit can also be studied.
The $W$ $p_T$ distribution is mainly the result of the 
intrinsic $p_T$ of the incoming partons and the initial state radiation. 
These largely universal mechanisms can be constrained with dilepton events.
The Drell-Yan continuum, between 20~GeV and $M_Z$, allows to measure the dilepton $p_T$ distribution in the 
$W$ mass range, providing a strong lever arm on the $W$ $p_T$. 
The $W$ rapidity distribution is essentially driven by the proton PDFs. 
At the LHC, $Z$ and $W$ bosons are essentially produced through sea quark interactions. 
We thus expect a strong correlation between the $W$ and $Z$  production. In particular,
a precise measurement of $d\sigma/d y (Z)$ will constrain the $W$ rapidity distribution.
The study~\cite{W_paper} shows that the uncertainty induced by strong interactions 
can be limited to 4~MeV with 10~fb$^{-1}$.

This study, more exhaustive than what can be described here, concludes that an overall sensitivity of about
7~MeV per decay channel and with 10~fb$^{-1}$ is a reasonable goal.

\section{Conclusion}

Recent studies by ATLAS and CMS collaborations show that uncertainties of the order of 
$\delta M_W\simeq 5$~MeV and $\delta M_t\simeq 1$~GeV are reachable, which translate into
$\delta M_H\simeq 15$~GeV. Beyond discovery, the LHC should allow to bring a deeper 
understanding of the electroweak symmetry breaking.

\begin{footnotesize}

\end{footnotesize}


\end{document}